\documentstyle[prb,aps,twocolumn,graphicx]{revtex}
\begin{document}
\draft

\twocolumn [\hsize\textwidth\columnwidth\hsize\csname
@twocolumnfalse\endcsname

\title{The properties of Haldane excitations and multi-particle
states\\ in the antiferromagnetic spin-1 chain compound ${\bf
CsNiCl_{3}}$}

\author{M. Kenzelmann,$^{1}$ R.~A. Cowley,$^{1}$ W.~J.~L. Buyers,$^{2,3}$
Z. Tun,$^{2}$ R. Coldea$^{4,5}$ and M. Enderle$^{6,7}$}

\address{(1) Oxford Physics, Clarendon Laboratory, Oxford OX1 3PU,
United Kingdom\\(2) Neutron Program for Materials Research,
National Research Council of Canada, Chalk River, Ontario, Canada
KOJ 1J0\\(3) Canadian Institute for Advanced Research, Toronto,
Ontario, Canada\\(4) Oak Ridge National Laboratory, Solid State
Division, Oak Ridge, Tennessee 37831\\(5) ISIS Facility,
Rutherford Appleton Laboratory, Oxon OX11 0QX, United Kingdom\\(6)
Technische Physik, Geb\"{a}ude 38,  Universit\"{a}t des
Saarlandes, 66123 Saarbr\"{u}cken, Germany\\(7) Institut
Laue-Langevin, BP 156 38042 Grenoble, Cedex 9, France}

\date{\today}
\maketitle

\begin{abstract}
We report inelastic time-of-flight and triple-axis neutron
scattering measurements of the excitation spectrum of the coupled
antiferromagnetic spin-1 Heisenberg chain system ${\rm
CsNiCl_{3}}$. Measurements over a wide range of wave-vector
transfers along the chain confirm that above $T_N$ ${\rm
CsNiCl_{3}}$ is in a quantum-disordered phase with an energy gap
in the excitation spectrum. The spin correlations fall off
exponentially with increasing distance with a correlation length
$\xi=4.0(2)$ sites at $T=6.2\;\mathrm{K}$. This is shorter than
the correlation length for an antiferromagnetic spin-1 Heisenberg
chain at this temperature, suggesting that the correlations
perpendicular to the chain direction and associated with the
interchain coupling lower the single-chain correlation length. A
multi-particle continuum is observed in the quantum-disordered
phase in the region in reciprocal space where antiferromagnetic
fluctuations are strongest, extending in energy up to twice the
maximum of the dispersion of the well-defined triplet excitations.
We show that the continuum satisfies the Hohenberg-Brinkman sum
rule. The dependence of the multi-particle continuum on the chain
wave-vector resembles that of the two-spinon continuum in
antiferromagnetic spin-1/2 Heisenberg chains. This suggests the
presence of spin-1/2 degrees of freedom in ${\rm CsNiCl_{3}}$ for
$T \leq 12\;\mathrm{K}$, possibly caused by multiply-frustrated
interchain interactions.
\end{abstract}

\pacs{PACS numbers: 75.25.+z, 75.10.Jm, 75.40.Gb}

]

\newpage

\section{Introduction}
Nearly two decades after Haldane's conjecture \cite{Haldane83} the
interest in quantum spin dynamics of low-dimensional
antiferromagnets continues unabated. Much of the research is
focused on one-dimensional (1D) antiferromagnets with small spin
$S$, where quantum fluctuations are strong because the classical
${\rm N\acute{e}el}$ ground state is not an eigenstate of the
Hamiltonian. The ground state of such an antiferromagnetic (AF)
chain is a spin singlet and long-range magnetic order is absent,
even at $T=0\;\mathrm{K}$. The spin correlation function in the
ground state of AF Heisenberg chains depends on the spin quantum
number $S$: for $S=1/2$ the spin-spin correlations fall off as a
power law with increasing distance (quasi long-range ordered
ground state) but for $S=1$ the spin-spin correlations fall off
exponentially with increasing distance.\cite{Haldane83} The
excitation spectrum is also different: for $S=1/2$ the elementary
excitations are spin-1/2 particles called spinons and the spectrum
extends to zero energy transfer while for $S=1$ the elementary
excitations are spin-1 particles with a gap (Haldane
excitations).\par

Inelastic neutron scattering is one of the most direct methods to
measure the magnetic excitations in solids. In neutron scattering
experiments, the spin of the neutron changes by $\Delta
S_z=0,\pm1$ along a quantization axis so that at low temperatures
the response is dominated by broad continua scattering if the
elementary particles carry spin-1/2. Sharp modes are observed,
however, if the elementary particles carry spin-1. The spectra
observed with neutron scattering are therefore fundamentally
different in AF spin-1/2 and spin-1 Heisenberg chains: for a
spin-1 chain at low temperatures the excitation spectrum is
dominated by well-defined modes\cite{Buyers86,Morra,Ma} while in
the case of a spin-1/2 chain the dominant contribution is a
continuum consisting of two-spinon states. The two-spinon
continuum has a characteristic wave-vector dependence reflecting
the momentum and energy addition of two particles and it has an
upper energy boundary at the AF point which is double the maximum
single-particle energy and decreases away from it. The observation
of such a scattering continuum in a spin system\cite{Tennant93}
thus clearly suggests that spin-1/2 particles are its elementary
excitations rather than spin-1 particles.\par

Considering the fundamental difference between AF spin-1/2 and
spin-1 chains, it came as a surprise when we observed in the
spin-1 chain compound ${\rm CsNiCl_3}$ a multi-particle scattering
continuum that resembles the two-spinon continuum of AF spin-1/2
chains.\cite{Kenzelmann_CsNiCl3_continuum} Its wave-vector
dependence and its intensity suggest that it does not originate
from two- or three-particle excitations of the elementary spin-1
excitations but instead from spin-1/2 degrees of freedom in the
system. Their detailed study promises to shed light on the
transition from spin-1/2 particles in quantum critical AF spin-1/2
chains and the spin-1 particles found in spin liquids like the
gapped Haldane quantum antiferromagnets.\par

The focus of our previous paper\cite{Kenzelmann_CsNiCl3_continuum}
was on the excitation spectrum of the quantum-disordered phase at
the AF point $Q_c=1$ and it was shown with three independent
neutron scattering measurements that there is a multi-particle
scattering continuum extending from the gapped onset of the
excitations up to $12\;\mathrm{meV}$. The present paper contains a
study of the excitation spectrum for a wide range of wave-vector
transfers and for both the well-defined excitations and the
multi-particle continuum. For the well-defined excitations the
results largely confirm previous results in the quantum-disordered
phase of ${\rm CsNiCl_3}$.\cite{Buyers86,Steiner,Zaliznyak} Data
measured with a sample mount which consists exclusively of
aluminum to avoid any background scattering originating from
hydrogenous materials confirm the previous estimate that $12(2)\%$
of the scattering at the AF point $Q_c=1$ is carried by the
multi-particle continuum. A detailed description of the
multi-particle states is given for a wide range of wave-vector
transfers. The integrated intensity of the scattering is shown to
be consistent with the first-moment sum rule. The characteristics
of the continuum are compared to those of the two-spinon continuum
in AF spin-1/2 Heisenberg chains.\par

\section{Properties of ${\bf CsNiCl_3}$}
\begin{figure}
  \includegraphics[height=4.8cm,bbllx=110,bblly=280,bburx=515,
  bbury=515,angle=0,clip=]{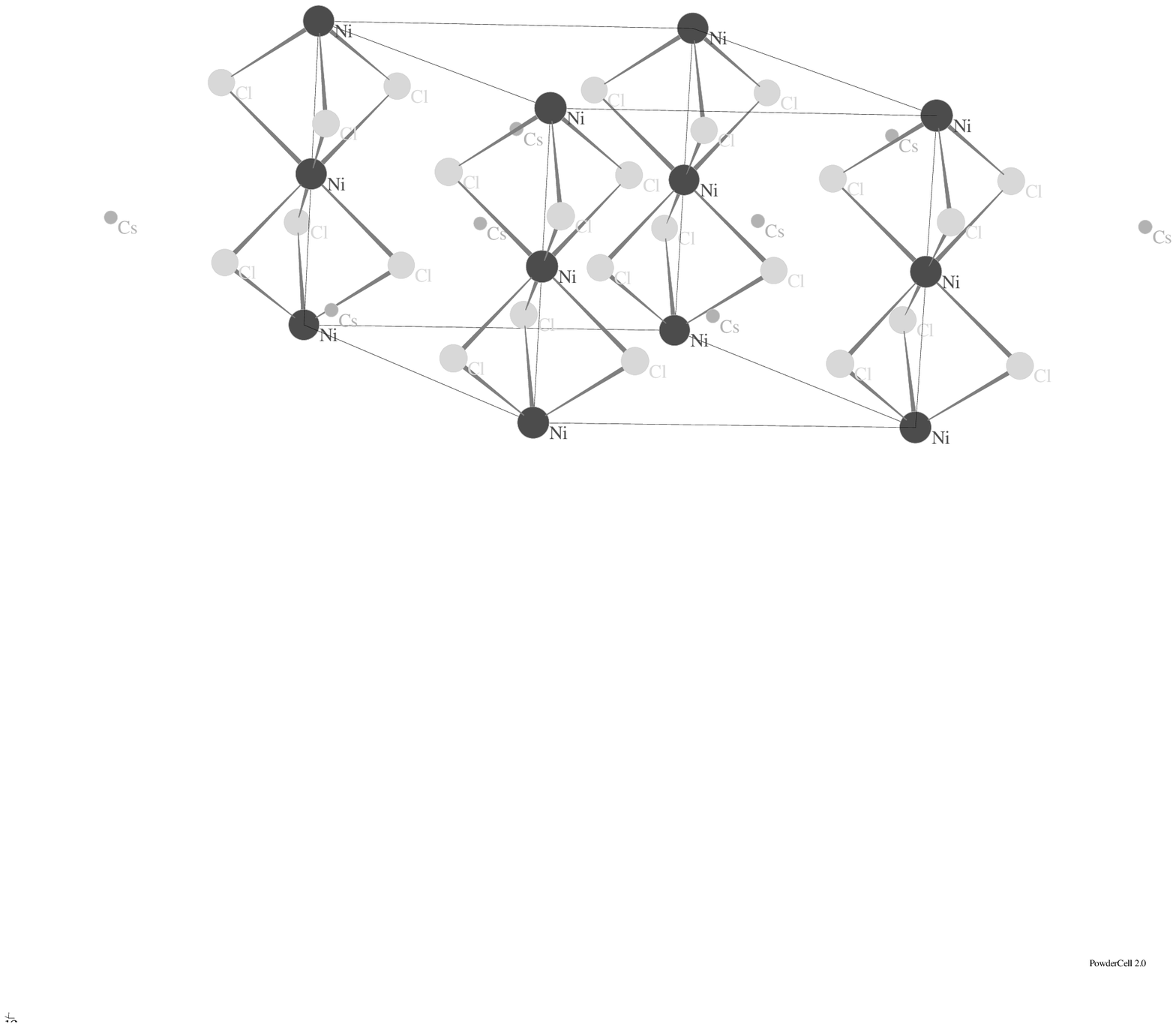}
  \vspace{0.2cm}
  \caption{Crystal structure of ${\rm CsNiCl_3}$. The Ni atoms are
  shown as dark spheres, and the Cs and Cl are shown as light
  spheres. The figure shows four Ni chains on a triangular lattice
  running along the c-axis. The Ni atoms are coupled along the c-axis
  by an AF superexchange interaction along the
  Ni-Cl-Ni paths.}
  \label{CsNiCl3-crystal}
\end{figure}

\begin{figure}
  \includegraphics[height=5cm,bbllx=100,bblly=210,bburx=450,
  bbury=460,angle=0,clip=]{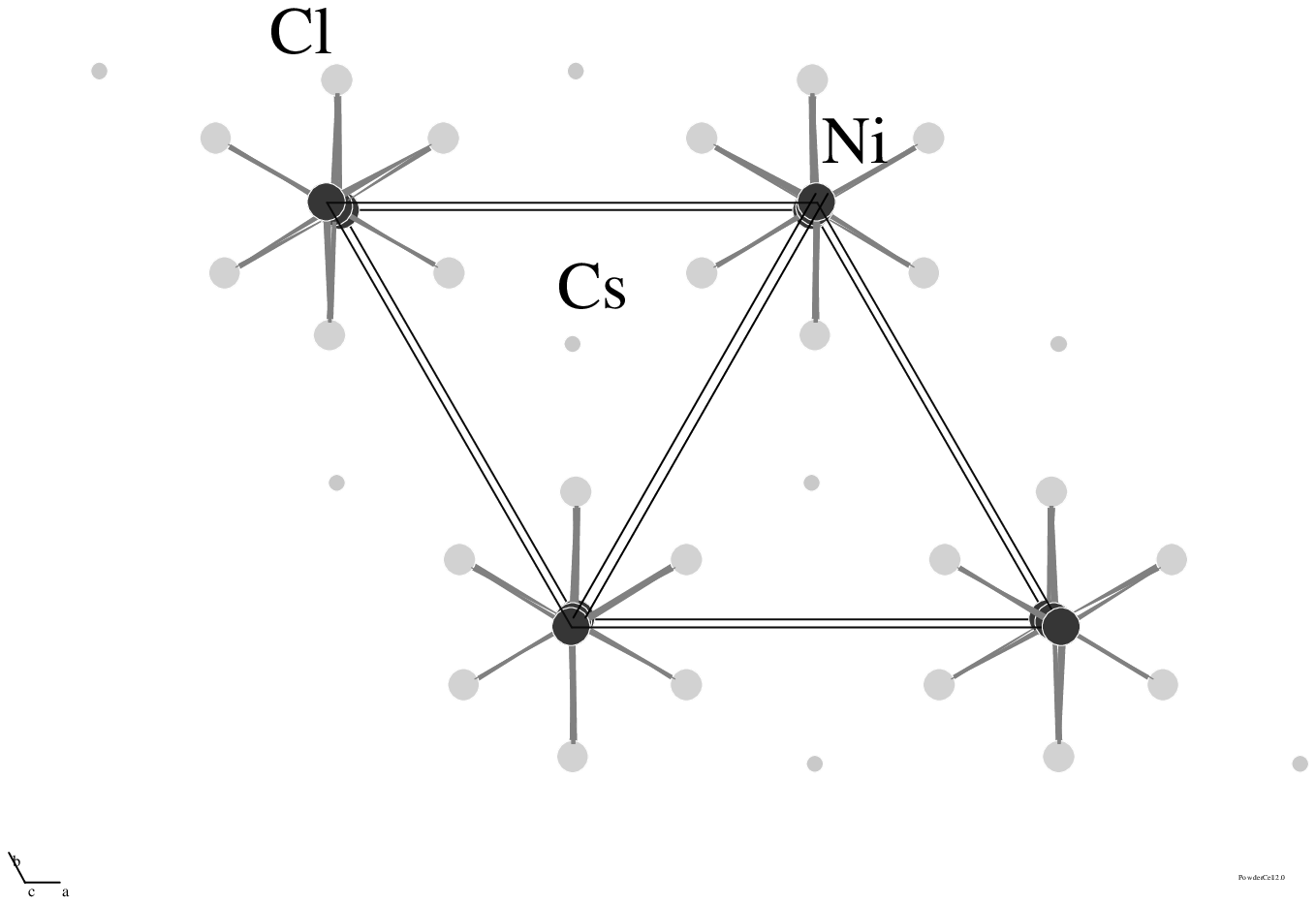}
  \vspace{0.2cm}
  \caption{Crystal structure of ${\rm CsNiCl_3}$ projected onto the
  basal plane. The Ni atoms are shown as dark spheres, and the Cs and
  Cl are shown as light spheres. Ni atoms belonging to different chains
  are coupled by an AF superexchange interaction along
  the Ni-Cl-Cl-Ni paths.}
  \label{CsNiCl3-crystal-basal}
\end{figure}

${\rm CsNiCl_{3}}$ is one of the most studied AF coupled spin-1
chain systems. It crystallizes in a hexagonal crystal structure,
$D^{4}_{6h}$ space group, and the lattice constants at low
temperatures are $a=7.14\,$\AA \, and $c=5.90\,$\AA \,
(Fig.~\ref{CsNiCl3-crystal}). The magnetic moments are carried by
${\rm Ni^{2+}}$-ions which interact via a superexchange
interaction involving ${\rm Cl^{-}}$-ions. Because the
superexchange path along the c-axis contains only one ${\rm
Cl^{-}}$-ion and perpendicular to the c-axis two ${\rm
Cl^{-}}$-ions (Figs.~\ref{CsNiCl3-crystal},
\ref{CsNiCl3-crystal-basal}), the superexchange between the ${\rm
Ni^{2+}}$-ions along the c-axis is much stronger than
perpendicular to the c-axis. The spin Hamiltonian of ${\rm
CsNiCl_{3}}$ is that of a system of coupled spin-1 chains with a
strong intrachain interaction $J$ and a weak interchain
interaction $J'$. It can be written as
\begin{equation}
{\mathcal{H}}=J \sum_{i}^{\rm chain} {\bf{S}}_i \cdot
{\bf{S}}_{i+1} + J' \sum_{<i,j>}^{\rm plane} {\bf{S}}_i \cdot
{\bf{S}}_j - D \sum_{i} (S_{\rm i}^{z})^2\, . \label{Hamiltonian}
\end{equation}

The superexchange interaction $J$ along the c-axis was measured
with a high-field magnetization measurement of the magnetic
saturation field and is $2.28\;\mathrm{meV}$.\cite{Katori} The
superexchange interaction perpendicular to the c-axis is
$J'=0.044\;\mathrm{meV}$ as determined from the measurement of the
spin-wave energies in the antiferromagnetically ordered structure
at $T=2\;\mathrm{K}$ and comparing them with spin-wave theory.
\cite{Buyers86,Morra} The weak easy-axis Ising anisotropy
$D=4\;\mathrm{\mu eV}$, which was derived from the spin-flop
transition field extrapolated to $T=0\;\mathrm{K}$, is small
enough that ${\rm CsNiCl_{3}}$ is a good example of an isotropic
Heisenberg antiferromagnet.\cite{Kadowaki}\par

Due to the interchain interaction, the coupled spin chain system
undergoes three-dimensional (3D) long-range ordering below
$T_N=4.84\;\mathrm{K}$, where the c-axis components of the
magnetic moments order.\cite{Kadowaki} At $T_{\rm canted}$=
$4.4\;\mathrm{K}$, the components perpendicular to the c-axis
order too. In the resulting magnetic structure, $\frac{2}{3}$ of
the spins are canted away from the c-axis by $44^{o}$ at
$4.4\;\mathrm{K}$ and $59^{o}$ at
$1.6\;\mathrm{K}$,\cite{Cox_Minkiewicz,Yelon} due to the easy-axis
anisotropy and the magnetic frustration of the triangular lattice.
Extrapolated to $T=0\;\mathrm{K}$, the ordered magnetic moment is
$1.05\mu_B$, considerably less than the free-ion value
$2\mu_B$.\cite{Minkiewicz} This reduction can be ascribed to
strong quantum fluctuations in the 3D ordered phase. Strong
quantum fluctuations in the 3D ordered phase are also apparent in
the excitation spectrum, where excitations derived from
longitudinal Haldane modes were observed which cannot be described
by conventional spin-wave theory.\cite{Enderle99} Their existence
is related to longitudinal fluctuations of the ordered moment, as
shown in a calculation of the excitations which includes the
gapped Haldane mode coupled to transverse gapless spin-wave-like
excitations.\cite{Affleck_Wellman}\par

In the quantum-disordered phase above $T_N$, the excitation
spectrum of ${\rm CsNiCl_{3}}$ has an energy gap (Haldane
gap)\cite{Buyers86} with a minimum at the AF wave-vector. The gap
has roughly the energy expected of a Haldane gap, which is a
consequence not of a single ion anisotropy but of strong quantum
fluctuations. The excitations are a triplet of well-defined $S=1$
modes,\cite{Steiner,Tun88} and the spin-spin static correlation
function falls off exponentially with increasing distance. For
wave-vectors near the nuclear zone center, a broadening of the
excitation spectrum may arise from the theoretically predicted
two-particle continuum.\cite{Affleck_Weston}\par

The magnetic susceptibility and the magnetic specific heat have a
broad maximum near $T=30\;\mathrm{K}$,\cite{Achiwa,Moses}, which
had been traditionally interpreted as the temperature where
short-range correlations set in. A neutron scattering experiment
showed that the excitations at the AF point persist as a resonance
up to at least $70\;\mathrm{K}$
\cite{Kenzelmann_CsNiCl3_gap,Kenzelmann_CsNiCl3_temperature}, so
that short-range correlations survive even to temperatures of the
order of the spin band width.\par

\section{Experimental Details}
An inelastic neutron scattering experiment measures the response
of a coupled spin system as a function of energy and momentum
transfer. The magnetic scattering cross section is directly
related to the dynamic structure factors $S^{\alpha
\beta}(Q,\omega)$, which are the Fourier transform of the space-
and time-dependent spin-pair correlation functions. Defining the
neutron energy and momentum transfers to the spin system as
$\omega=E_i-E_f$ and $\bbox{Q}=\mathbf{k}_i-\mathbf{k}_f$,
respectively, the magnetic scattering cross section for
unpolarized neutrons can be written as
\begin{equation}
\frac{d^2\sigma}{d\Omega d\omega}
=|f(\bbox{Q})|^2\frac{k_f}{k_i}\sum_{\alpha,\beta}(\delta_{\alpha
\beta}-\hat{Q}_{\alpha} \hat{Q}_{\beta})S^{\alpha
\beta}(\bbox{Q},\omega)\, ,
\end{equation}where $\alpha,\beta=x,y,z$ and $f(\bbox{Q})$ is
the magnetic form factor. For a spin system with Heisenberg
interactions in the paramagnetic phase, the entire magnetic
response is inelastic and the dynamic structure factors have the
property $S(Q,\omega)=S^{xx}$= $S^{yy}=S^{zz}$ because the system
has unbroken rotational symmetry. For the Hamiltonian of ${\rm
CsNiCl_{3}}$, the total z-component of spin is a constant of
motion if $D \sum_{i} (S_{\rm i}^{z})^2$ is neglected ($S^z=\sum_i
S_i^z$ commutes with the Hamiltonian) and $S^{\alpha \beta}=0$ for
$\alpha \neq \beta$.\par

A single crystal of ${\rm CsNiCl_{3}}$ $5\; \times 5\; \times
20\;\mathrm{mm^3}$ was used for the experiments. The experiments
were performed using the triple-axis spectrometer DUALSPEC at the
Chalk River Laboratories, Canada, and the chopper time-of-flight
spectrometer MARI at the pulsed spallation source ISIS of the
Rutherford Appleton Laboratory, United Kingdom. The sample was
kept sealed in an aluminum can containing helium gas to prevent
the absorption of water. For the DUALSPEC experiment, the sample
was wired to an aluminum plate and placed inside an aluminum can
such that no sealant could cause extraneous scattering of
neutrons.\par

The measurements with the DUALSPEC triple-axis spectrometer were
performed at $6\;\mathrm{K}$. The sample was mounted in a cryostat
with its (hhl) crystallographic plane in the horizontal scattering
plane. The scattered neutron energy was set to
$14.51\;\mathrm{meV}$. A graphite filter was used to absorb the
higher order reflections from a vertically focusing pyrolytic
graphite monochromator and a flat analyzer. The effective
collimation from reactor to detector was $39'$-$48'$-$51'$-$72'$
and gave an energy resolution of $0.8\;\mathrm{meV}$, as
determined from the quasi-elastic incoherent peak. The
longitudinal wave-vector width of the (002)-peak and (220)-peak
were $0.022$ and $0.019$ (full width at half maximum (FWHM) in
reciprocal lattice units), respectively, and the calculated
vertical resolution was $0.29$\,\AA$^{-1}$.

\begin{figure}
  \includegraphics[height=5cm,bbllx=85,bblly=301,bburx=504,
  bbury=585,angle=0,clip=]{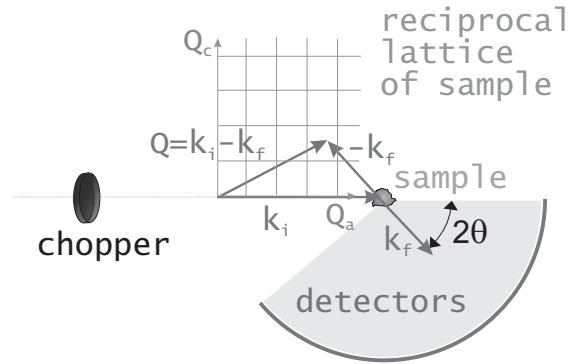}
  \vspace{0.2cm}
  \caption{Schematic drawing of the chopper time-of-flight
  spectrometer MARI at ISIS. A monochromatic neutron beam
  is generated using a chopper. The neutrons are scattered
  at the sample and then detected in the detector banks
  located at a distance of $4.02\;\mathrm{m}$ in a half circle
  below the sample. Depending on the incident energy and the
  orientation of the sample, different wave-vector and energy
  transfer regions are accessed during an experiment.}
  \label{MARI-time-of-flight}
\end{figure}

The time-of-flight spectrometer MARI produced a monochromatic beam
by means of the Fermi chopper spinning at $150\;\mathrm{Hz}$. The
scattered neutrons were detected by an array of ${\rm He^{3}}$
tube detectors typically $30\;\mathrm{cm}$ long (perpendicular to
the principal scattering plane) and with a diameter of
$2.5\;\mathrm{cm}$ situated $4.02\;\mathrm{m}$ from the sample
position. The low-angle bank consists of $136$ detectors arranged
symmetrically around the incident neutron beam direction and
having scattering angles from $3.43$ to $13.29\mathrm{^{o}}$. The
detectors in the high-angle banks are arranged in three strips
vertically below the sample and cover the scattering angles from
$12.43$ to $134.14\mathrm{^{o}}$, where the detector tubes are
aligned perpendicular to the vertical scattering plane. The
central strip is in the vertical scattering plane, and the two
side strips of detectors are out of the vertical plane by an
average of $5.56\mathrm{^{o}}$ for high scattering angles. At
small scattering angles, the detectors in the two side strips are
out of the principal scattering plane by up to
$21.36\mathrm{^{o}}$ and the neutrons detected in these detectors
have a considerable wave-vector component perpendicular to the
scattering plane.\par

The experiment using MARI was performed with the sample oriented
such that the (hhl) crystallographic plane was in the vertical
scattering plane and the c-axis was perpendicular to the incident
beam direction. The incident neutron energy $E_i$ was set either
to $20$ or $30\;\mathrm{meV}$ and the energy resolution was $0.35$
and $0.4\;\mathrm{meV}$, respectively, as determined from FWHM of
the quasi-elastic incoherent scattering peak. The resolution in
wave-vector transfer at zero energy transfer was typically
$0.02$\,\AA$^{-1}$ along the $c^{\star}$-axis and along the
$[110]$ direction and up to $0.19$\,\AA$^{-1}$ perpendicular to
the scattering plane if only the central detector bank was used.
Both the energy and the wave-vector resolution improved with
increasing energy transfer. The measurements were performed with
the sample at temperatures between $6.2$ and $12\;\mathrm{K}$. The
scattering was measured for a total proton charge between $4000$
and $8600\;\mathrm{\mu Ahr}$ at an average proton current of
$~170\;\mathrm{\mu A}$.\par

The simultaneous measurement of continuous energy transfers
$\omega$ together with the large range of scattering angles allows
the measurement of the dynamical structure factor
$S({\bf{Q}},\omega)$ over the scattering surface, which is a
two-dimensional subspace of ($Q_a$, $Q_c$, $\omega$), where both
$Q_a$ and $Q_c$ lie in the scattering plane. The scattering
surface may then be projected down onto ($Q_c$, $\omega$)-plane
and scans with a constant energy transfer or a wave-vector
transfer along the chain can be obtained by binning the data
appropriately. These scans will henceforth be called
constant-$\omega$ and constant-$Q_c$ scans, respectively. In our
case, because ${\rm CsNiCl_{3}}$ is a 1D magnet and the interchain
dispersion is sufficiently small, $S(Q_c,\omega)$ mainly reflects
the excitation spectrum of the spin chains.\par

\section{Experimental results}
\subsection{Haldane excitations}
\begin{figure}
\begin{center}
  \includegraphics[height=9cm,bbllx=50,bblly=155,bburx=520,
  bbury=640,angle=0,clip=]{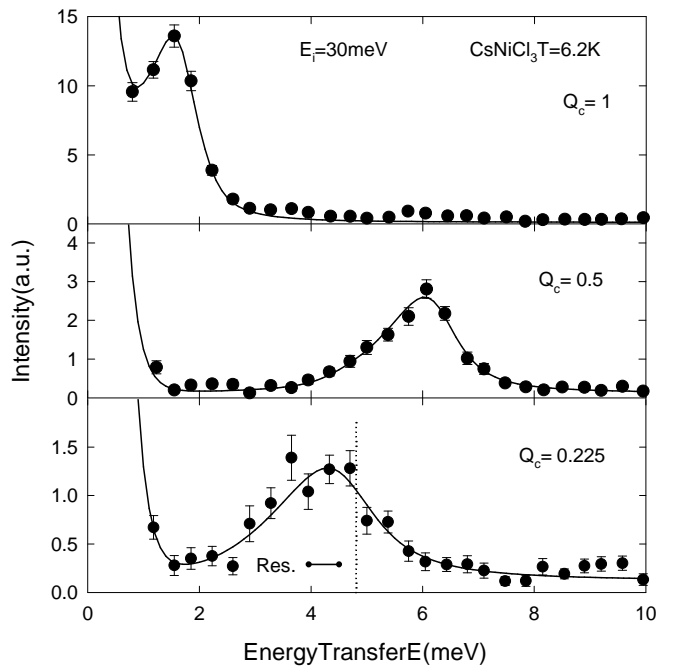}
  \vspace{0.2cm}
  \caption{Neutron scattering intensity at $6.2\;\mathrm{K}$
  as a function of energy transfer for three different
  wave-vector transfers $Q_c$ along the chain. The data
  were measured using the central detector bank of the MARI
  time-of-flight spectrometer and the wave-vector range sampled
  for each $Q_c$ in $\pm 0.025$. The incident energy $E_i$
  was $30\;\mathrm{meV}$. The solid line is a fit as described
  in the text and the dotted line shows the onset of the
  two-magnon continuum as predicted by the ${\rm NL\sigma M}$.}
  \label{scans-6K}
\end{center}
\end{figure}

The quantum-disordered magnetic phase of ${\rm CsNiCl_3}$ at
temperatures between $6.2$ and $12\;\mathrm{K}$ was studied for a
wide wave-vector and energy range using the MARI time-of-flight
spectrometer. Fig.~\ref{scans-6K} shows three constant-$Q_c$ scans
at $6.2\;\mathrm{K}$ obtained using the binning procedure
described above. Only data measured in the central detector bank
are shown in the Fig.~\ref{scans-6K}. This is because at low
scattering angles the two side detector banks are, as mentioned,
considerably out of the vertical scattering plane and so the
dynamic structure factor $S(\bbox{Q},\omega)$ is different for the
three detector banks.\par

The data are treated as if there was a well-defined excitation and
some extra scattering\cite{Kenzelmann_CsNiCl3_continuum} but the
data could also be interpreted as a continuum with a well-defined
onset at the low-energy boundary. The well-defined excitation has
the most intensity at the AF point $Q_c=1$, which corresponds to a
$\pi$ phase difference between the Ni spins along the chains. The
intensity decreases towards $Q_c=0.5$, but it is visible down to
$Q_c=0.2$. The scattering was fitted with an antisymmetrized
Lorentzian form weighted by the Bose factor
\begin{eqnarray}
        S(\bbox{Q},\omega)= A \cdot \left( n(\omega)+1 \right)
        \cdot \hspace{1cm} \nonumber \\
        \left(\frac{\Gamma}{(\omega-\epsilon(\bbox{Q}))^2 +\Gamma^2} -
        \frac{\Gamma}{(\omega+\epsilon(\bbox{Q}))^2+\Gamma^2}\right)\, .
        \label{Lorentzian form}
\end{eqnarray}
and convoluted with the line-shape of the quasi-elastic incoherent
scattering, which is a good estimate of the resolution
particularly at low energy transfers. Here, $\epsilon(\bbox{Q})$
is the dispersion relation, $\Gamma$ is the excitation width,
$n(\omega)+1$ is the population factor, $\omega$ is the neutron
energy transfer and $A$ is a scaling constant. For the excitations
close to $Q_c=1$, we found that a Lorentzian line-shape gave fits
with consistently lower sum of discrepancies $\chi^2$ than for an
antisymmetrized Gaussian cross-section, in agreement with our
measurements obtained using the RITA spectrometer
\cite{Kenzelmann_CsNiCl3_gap,Kenzelmann_CsNiCl3_temperature}.\par

\begin{figure}
\begin{center}
  \includegraphics[height=6.4cm,bbllx=80,bblly=265,bburx=485,
  bbury=570,angle=0,clip=]{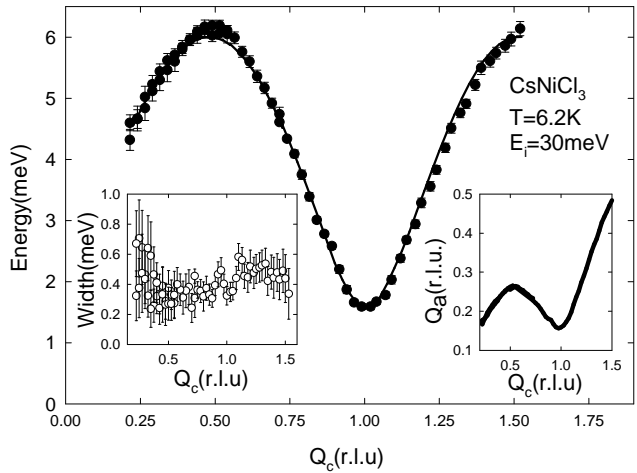}
  \vspace{0.2cm}
  \caption{Excitation energy as a function of wave-vector
  transfer $Q_c$ along the chain. The solid line is a fit
  to the excitation energy as described in the text. Left inset:
  Lorentzian half width $\Gamma$ as a function of wave-vector transfer
  $Q_c$ along the chain. Right inset: Wave-vector transfer
  along $[Q_a,Q_a,0]$ and $Q_c$ calculated from the fitted
  energy transfer and $Q_c$.}
  \label{dispersion-width-6K}
\end{center}
\end{figure}

Fits were performed to obtain the intensity, the frequency and the
width of the well-defined excitations. An energy-independent
background increasing with the wave-vector transfer $Q_c$ along
the chain was chosen such that the scattering at high energy
transfers matches the scattering at low energies below the
well-defined excitation. The excitation energy and width are shown
in Fig.~\ref{dispersion-width-6K}. The data for $Q_c > 1.6$ were
not analyzed due to their contamination with phonons. The maximum
of the dispersion is $6.19(9)\;\mathrm{meV}$, in agreement with
previous measurements \cite{Morra} and the excitation energy at
$Q_c=1$ is $1.59(5)\;\mathrm{meV}$.\par

The observed dispersion of a single AF spin-1 chain is often
written phenomenologically as \cite{Ma}
\begin{equation}
     \omega(Q_c) = \sqrt{ \Delta^2 + v_{\rm s}^2
     \sin^2(Q_c \pi)+\alpha^2 \cos^2(\frac{Q_c}{2} \pi)}
     \label{1D-dispersion}\, ,
\end{equation}
where $\Delta$ is the Haldane gap, $v_{\rm s}$ is the spin
velocity and $\alpha$ determines the asymmetry of the dispersion
about $Q_c=0.5$. For a two-particle gap of $2\Delta$ as
$Q_c\rightarrow0$ one would expect $\alpha=\sqrt{3}\Delta$.
Because the chains are coupled in ${\rm CsNiCl_3}$ the excitations
have a dispersion perpendicular to the chain direction. The
coupling of the chains can be taken into account in a Random Phase
Approximation (RPA) and the dispersion relation can be written as
\cite{Tun90}
\begin{eqnarray}
     &\omega^{2}(\bbox{Q}) = \omega(Q_c)
     \cdot \left( \omega(Q_c)+ 2 \cdot J'(Q_a,Q_b)
     \cdot S(\bbox{Q}) \right)\, .&
     \label{normalized_dispersion}
\end{eqnarray}$S(\bbox{Q})$ is the static structure factor
and $J'(Q_a,Q_b)$ is the Fourier transform of the exchange
interaction perpendicular to the chain axis. The Fourier transform
$J'(Q_a,Q_a)$ along the reciprocal $[1,1,0]$-direction, which is
of relevance for the present experiment, is given by
\begin{equation}
     J'(Q_a,Q_a) = J' \cdot (4 \cos(2\pi Q_a)+ 2\cos(4\pi Q_a) )\,
     ,
     \label{Fourier_J}
\end{equation}and vanishes for $Q_a=0.19$. The static structure
factor falls rapidly as $|Q_c-1|$ increases, weakening the effect
of the interchain exchange interaction on the dispersion.\par

The solid line in Fig.~\ref{dispersion-width-6K} corresponds to a
fit of Eq.~\ref{normalized_dispersion} to the observed dispersion
taking into account the wave-vector dependence of the measured
excitation perpendicular to the chain (inset of
Fig.~\ref{dispersion-width-6K}) and fixing the excitation band
width perpendicular to the chain axis by setting
$\omega(\frac{2}{3},\frac{2}{3},1) =
0.4\;\mathrm{meV}$.\cite{Steiner} The fitted parameters are
$\Delta=1.24(4)\;\mathrm{meV}=0.54(2)J$, $v_{\rm
s}=5.70(7)\;\mathrm{meV}=2.50(3)J$ and $\alpha=2.5(3)\;{\mathrm
meV}=1.1(1)J=2.0\Delta$.\par

$\Delta$ is consistent with our previous measurements
\cite{Kenzelmann_CsNiCl3_gap} and is higher than the gap energy of
$1.14\;\mathrm{meV}$ calculated for the non-linear sigma
model\cite{Jolicoeur_Golinelli} (${\rm NL\sigma M}$) which
includes the temperature renormalization at $6.2\;\mathrm{K}$.
This is possibly due to an additional upward renormalization of
the gap energy caused by the coupling of the spin
chains.\cite{Kenzelmann_CsNiCl3_gap} The observed energy at
$Q_c=1$, $\omega=1.59(5)\;\mathrm{meV}$, is higher than $\Delta$
because it corresponds to an excitation with a different
wave-vector transfer perpendicular to the chain axis. For
weakly-coupled spin chains $\Delta$ can be directly measured at
$\bbox{Q}=(0.19,0.19,1)$ where the Fourier transform of the
interchain coupling vanishes. In this experiment, however, the gap
excitation could only be sampled at $\bbox{Q}=(0.16,0.16,1)$ as
shown in the right inset of Fig.~\ref{dispersion-width-6K}, and
here the excitation energy is higher.\par

\begin{figure}
\begin{center}
  \includegraphics[height=6.4cm,bbllx=80,bblly=265,bburx=485,
  bbury=570,angle=0,clip=]{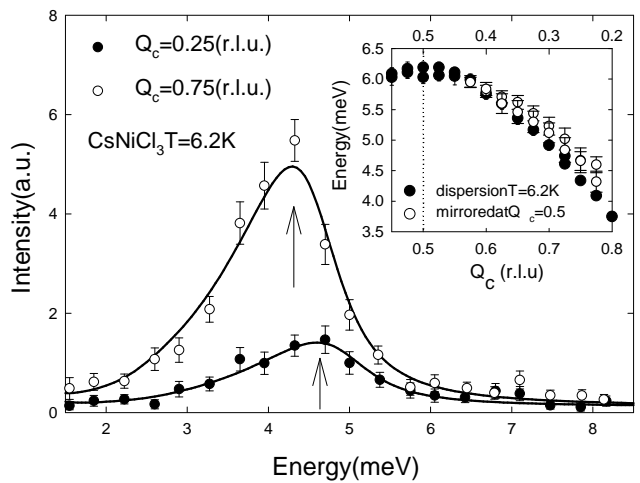}
  \vspace{0.2cm}
  \caption{Neutron scattering intensity as a function of energy
  transfer for $Q_c=0.25$ and $Q_c=0.75$. The energy of the
  excitation at $Q_c=0.25$ is clearly higher than at $Q_c=0.75$.
  Inset: The solid circles represent the excitation energy for
  $Q_c>0.5$ (bottom axis) and the open circles data for $Q_c<0.5$
  (top axis), both corrected for the $Q_a$-dependence of the
  dispersion. Hence the dispersion is not periodic
  about $Q_c=0.5$.}
  \label{2PI-period}
\end{center}
\end{figure}

The spin velocity $v_s$ is in excellent agreement with numerical
results predicting $v_s$ = $2.5J$.\cite{Sorensen_1,Sorensen_2}
$\alpha$ is consistent with previous results obtained for NENP
\cite{Ma} and ${\rm CsNiCl_3}$\cite{Zaliznyak_continuum} and its
non-zero value reflects the asymmetry of the dispersion with
respect to reflection at $Q_c=0.5$ as would occur for a simple
antiferromagnet. Fig.~\ref{2PI-period} shows the asymmetry about
$Q_c=0.5$ as measured using the MARI spectrometer. The figure
compares two equivalent scans at $Q_c=0.25$ and $Q_c=0.75$, and
the excitation at $Q_c=0.75$ clearly has a lower energy than the
one at $Q_c=0.25$. Due to the dependence of the excitations on
$Q_a$ (right inset of Fig.~\ref{dispersion-width-6K}), the energy
difference between the two observed excitations is increased
compared with what would have been observed for an isolated spin-1
chain. We applied the correction for this $Q_a$-dependence of the
dispersion by using Eq.~\ref{normalized_dispersion} and the
$Q_a$-dependence of the excitations (inset of
Fig.~\ref{dispersion-width-6K}) to extract the dispersion of an
uncoupled chain. The 3D corrected dispersion is shown in the inset
of Fig.~\ref{2PI-period} showing that the energy of the excitation
towards $Q_c=0$ is clearly higher than towards $Q_c=1$.\par


The $Q_c$-dependence of the excitation width
(Fig.~\ref{dispersion-width-6K}) suggests a cross-over from a
one-particle dispersion for $Q_c>0.5$ to a two-particle continuum
with the corresponding increased excitation width for $Q_c<0.5$.
The cross-over from a one-particle to a two-particle response has
also been observed by Zaliznyak \textit{et al.}
\cite{Zaliznyak_continuum} at $1.5\;\mathrm{K}$ in the 3D ordered
phase. In their experiment the increase in width is clearer
because the thermal broadening of the one-particle peaks is much
less. In both experiments the broadened peaks are centered at
energies lower than the onset of the two-particle continuum for
non-interacting chains (Fig.~\ref{dispersion-width-6K}). Possibly
this is because of the 3D inter-chain coupling which at $1.5$ and
$6.2\;\mathrm{K}$ gives considerable dispersion to the excitation
energies for $Q_c=1$ as $Q_a$ varies. This could decrease the
two-particle onset energy.\par

\subsection{Continuum scattering}
\begin{figure}
\begin{center}
  \includegraphics[height=6.4cm,bbllx=70,bblly=268,bburx=485,
  bbury=570,angle=0,clip=]{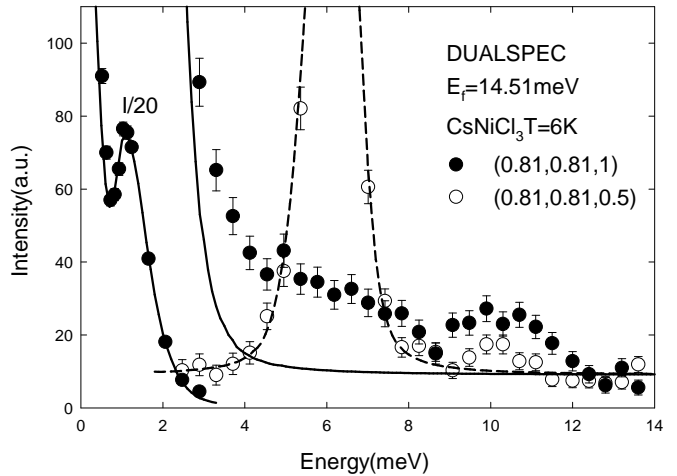}
  \vspace{0.2cm}
  \caption{Neutron scattering intensity at $\bbox{Q}=(0.81,0.81,1)$
  (solid circles) as a function of energy transfer measured using
  the DUALSPEC triple-axis spectrometer. The scattering intensity
  below $3\;\mathrm{meV}$ is shown 20 times reduced. The scattering
  at $\bbox{Q}=(0.81,0.81,0.5)$ is shown for comparison (open circles).
  The solid and dashed line are fits to the measured spectra as
  explained in the text.}
  \label{continuumQ-C5-highstatistics}
\end{center}
\end{figure}

Fig.~\ref{continuumQ-C5-highstatistics} shows two
constant-$\bbox{Q}$ scans measured using the DUALSPEC triple-axis
spectrometer at the Chalk River Laboratories and compares the
scattering at the AF point $\bbox{Q}=(0.81,0.81,1)$ with that at
$\bbox{Q}=(0.81,0.81,0.5)$, both at $T=6\;\mathrm{K}$. For $Q_c=1$
new data with the sample mounted so that no hydrogenous material
was in the beam are presented. The data are corrected for neutron
absorption. This was done by measuring the quasi-elastic
scattering when the sample presented the same angles to the
incident and scattered beams as in the constant-$\bbox{Q}$ scans.
At $Q_c=0.5$, where the dispersion of the well-defined excitation
has a maximum, the scattering intensity below and above the sharp
excitation is considerably lower than the intensity at $Q_c=1$,
and it gives a good measure of the nonmagnetic background. The
excess scattering at $Q_c=1$ at energies above the single-particle
excitation is the continuum scattering.\par

The well-defined excitation at $\bbox{Q}=(0.81,0.81,1)$ is well
described by an antisymmetrized Lorentzian weighted by the Bose
factor and convoluted with the resolution ellipsoid given by
Cooper and Nathans,\cite{Cooper_Nathans} but it does not account
for the continuum scattering. After subtracting phonon scattering
near $10\;\mathrm{meV}$, which was estimated from measurements at
large wave-vectors, the integrated continuum intensity is $9(2)\%$
of the total intensity at $\bbox{Q}=(0.81,0.81,1)$, consistent
with the previous estimate of
$12(2)\%$.\cite{Kenzelmann_CsNiCl3_continuum}\par

\begin{figure}
\begin{center}
  \includegraphics[height=9cm,bbllx=65,bblly=135,bburx=480,
  bbury=565,angle=0,clip=]{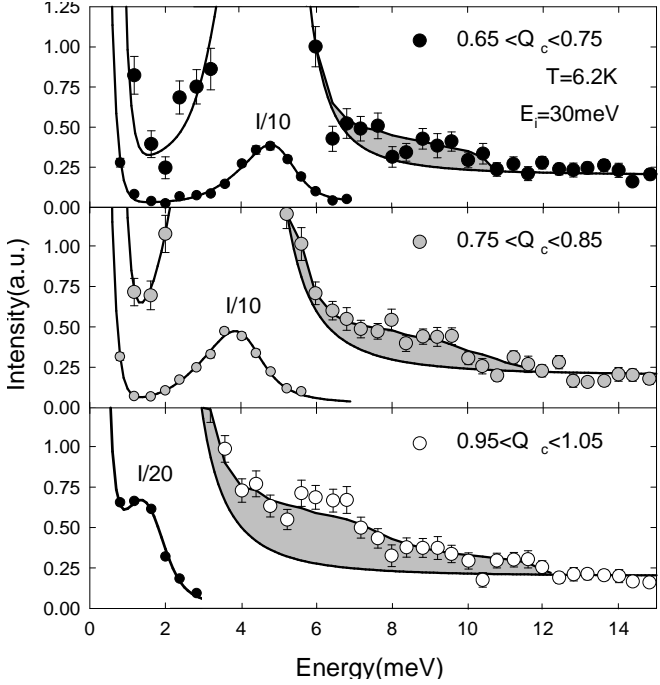}
  \vspace{0.2cm}
  \caption{Neutron scattering spectrum at $6.2\;\mathrm{K}$ for
  various wave-vector transfers. The peak intensities are scaled
  to fit onto the graph. The lower solid lines are fits to the
  peak data and the upper line is a guide to the eye enclosing
  the shaded area of the continuum scattering. The data were
  obtained on the MARI spectrometer with
  $E_i=30\;\mathrm{meV}$.}
  \label{continuumQ-cuts}
\end{center}
\end{figure}

The wave-vector dependence of the multi-particle continuum on
$Q_c$ was measured at $6.2\;\mathrm{K}$ using the time-of-flight
spectrometer MARI. Three constant-$Q_c$ scans at $\langle Q_c
\rangle=0.7$, $\langle Q_c \rangle=0.8$ and $\langle Q_c
\rangle=1$ shown in Fig.~\ref{continuumQ-cuts} confirm that for
energies higher than the energy of the well-defined excitation, a
slowly decreasing intensity is observed for all three wave-vector
transfers.\par

The observed spectra were fitted by an antisymmetrized Voigt
function weighted by the Bose factor and convoluted with the
line-shape of the quasi-elastic incoherent scattering peak, which
should give a good account of the energy resolution at low energy
transfers. A Voigt function, which is a convolution of a
Lorentzian with a Gaussian function, is needed to account for the
extra energy width resulting from the wide wave-vector resolution
$\Delta Q_c=0.1$. The fits give a good account of the well-defined
excitations (Fig.~\ref{continuumQ-cuts}), but do not account for
the scattering at higher energies associated with the
multi-particle continuum. For $\langle Q_c \rangle=1$, the
scattering extends up to about $12\;\mathrm{meV}$. The scattering
at $\langle Q_c \rangle=0.8$ and $\langle Q_c \rangle=0.7$ extends
to slightly lower energy transfers, indicating a decreasing upper
energy boundary of the continuum with increasing $|Q_c-1|$.\par

In order to confirm that the extra scattering is not a resolution
artifact, a series of tests using a spectrometer simulation
programme available at ISIS were made to investigate whether the
continuum above the well-defined excitations could result from the
convolution of the neutron resolution function with the single
mode dispersion. The program takes into account the spectrometer
parameters - the detailed pulse line-shape, the chopper
characteristics, the position and dimensions of the detectors etc.
- and predicts the scattering line-shape for a particular sample
orientation and model cross-section. The results of this
simulation reproduced the measured line-shape of the quasi-elastic
incoherent scattering and of the well-defined excitations, and
showed that none of the scattering at higher energies could arise
from the tail of the resolution function.\par

\begin{figure}
\begin{center}
  \includegraphics[height=9cm,bbllx=68,bblly=230,bburx=466,
  bbury=700,angle=0,clip=]{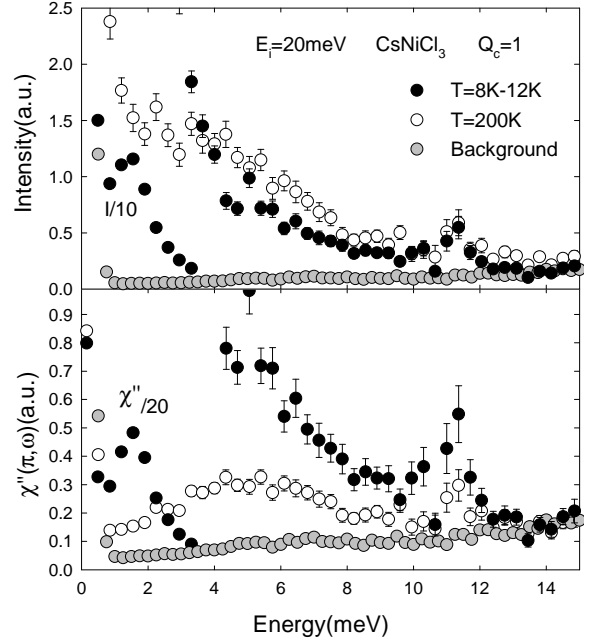}
  \vspace{0.1cm}
  \caption{Upper panel: Neutron scattering spectrum at $Q_c=1$, at
  the average of $T=8$ and $12\;\mathrm{K}$ (solid circles), and at
  $T=200\;\mathrm{K}$ (open circles). The intensity was integrated
  for $0.9 < Q_c < 1.1$. The data were measured using the MARI
  spectrometer and $E_i=20\;\mathrm{meV}$. Lower panel: the
  imaginary part of the magnetic susceptibility $\chi''(\pi,\omega)$
  as a function of energy transfer for the quantum-disordered phase
  at low temperatures and in the high-temperature limit. The grey
  solid circles give the estimated background as described in the
  text.}
  \label{Fig-background}
\end{center}
\end{figure}

We also investigated whether the scattering continuum can be
influenced by neutron scattering by phonons. Measurements were
performed at $8$, $12$ and $200\;\mathrm{K}$, with an incident
energy of $E_i=20\;\mathrm{meV}$, and with the 1D axis oriented
perpendicular to the incident neutron beam direction. The
scattering intensity, normalized to the incident beam monitor, is
shown in Fig.~\ref{Fig-background} at different temperatures.\par

\begin{figure}
\begin{center}
  \includegraphics[height=6.4cm,bbllx=70,bblly=260,bburx=480,
  bbury=570,angle=0,clip=]{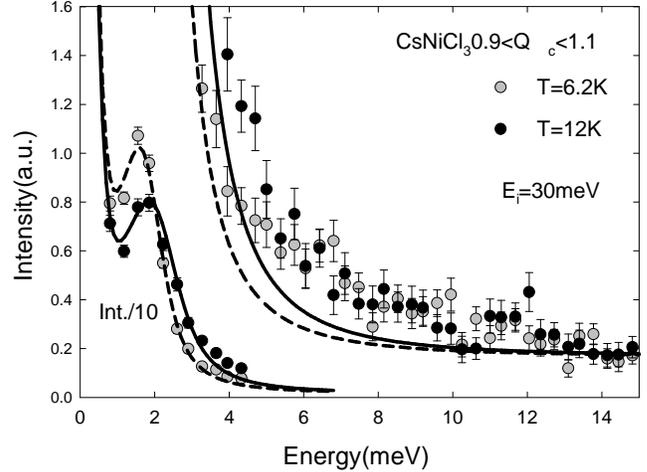}
  \vspace{0.1cm}
  \caption{Neutron scattering intensity as a function of energy
  transfer for wave-vector transfer with $0.9 < Q_c < 1.1$ at $6.2$
  and $12\;\mathrm{K}$. The peak from the Haldane gap mode is shown
  with its intensity reduced by a factor of 10. The lines are
  fits of a Lorentzian function to that observed intensity at
  $6\;\mathrm{K}$ (dashed line) and $12\;\mathrm{K}$ (solid line).}
  \label{continuum-6K-12K}
\end{center}
\end{figure}

Above $5\;\mathrm{meV}$, the intensity of the continuum scattering
does not change between $6$ and $12\;\mathrm{K}$
(Fig.~\ref{continuum-6K-12K}). We then averaged the spectra at $8$
and $12\;\mathrm{K}$ to obtain better statistics at high energy
transfers for $\langle T \rangle=10\;\mathrm{K}$. This leads to an
artificial broadening of the well-defined excitations, and so the
averaged data at low energy transfers were not analyzed in
detail.\par

The upper panel in Fig.~\ref{Fig-background} compares the neutron
scattering intensity at $Q_c=1$ measured with MARI in the quantum
disordered phase at $\langle T \rangle=10\;\mathrm{K}$ ($8 -
12\;\mathrm{K}$) with that measured at $200\;\mathrm{K}$. The
lower panel of Fig.~\ref{Fig-background} shows the imaginary part
of the generalized susceptibility $\chi''(Q_c=1,\omega)$ =
$S(Q_c=1,\omega)/(n(\omega)+1)$ obtained from the dynamic
structure factor. Yet at $200\;\mathrm{K}$ $\chi''(Q_c=1,\omega)$
is clearly less than at $\langle T \rangle=10\;\mathrm{K}$ for all
energy transfers, apart from a sharp peak at $11.5\;\mathrm{meV}$
which can be associated with a single-phonon excitation. It is not
present in the measurements with incident energy
$E_i=30\;\mathrm{meV}$ which sampled a different scattering
surface. If the continuum scattering between $5$ and
$12\;\mathrm{meV}$ for $\langle T \rangle=10\;\mathrm{K}$ were due
to neutron scattering by phonons, $\chi''(\pi,\omega)$ would be
comparable at $200\;\mathrm{K}$. The fact that it decreases
greatly shows that the slowly decreasing scattering intensity
above the Haldane mode is magnetic and not phonon scattering. At
$T=200\;\mathrm{K}$, $\chi''(Q_c=1,\omega)$ has a broad peak at
$\sim 5\;\mathrm{meV}$, as we will discuss in a forthcoming
publication.\cite{Kenzelmann_CsNiCl3_temperature} This broad peak
is absent at low temperatures when only low-energy magnetic
excitations are thermally activated and $\chi''(\pi,\omega)$ is
dominated by the slightly-broadened Haldane gap mode.\par

\begin{figure}
\begin{center}
  \includegraphics[height=6.3cm,bbllx=50,bblly=220,bburx=510,
  bbury=560,angle=0,clip=]{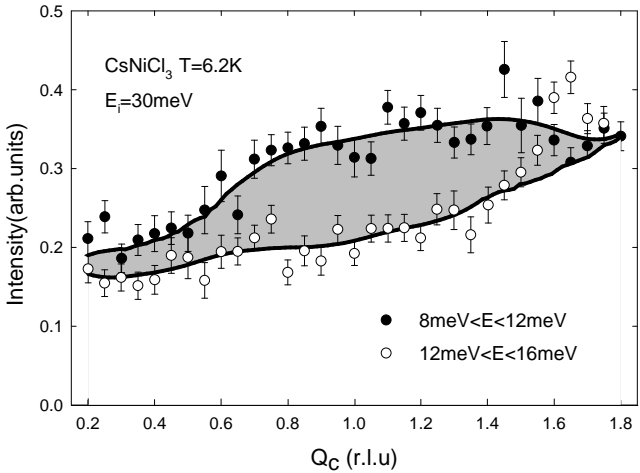}
  \vspace{0.1cm}
  \caption{Neutron scattering intensity at $T=6.2\;\mathrm{K}$ as
  a function of wave-vector transfer along the chain, $Q_c$, and
  integrated between $8$ and $12\;\mathrm{meV}$ (solid circles) and
  integrated between $12$ and $16\;\mathrm{meV}$ (open circles).
  Lines are guides to the eye. The data were measured in all three
  detector banks of the MARI spectrometer with an incident energy
  $E_i=30\;\mathrm{meV}$.}
  \label{continuum-cut}
\end{center}
\end{figure}

An estimate of the non-magnetic background was also obtained from
the scattering observed in detectors at high scattering angles
$2\Theta$ for which the wave-vector transfer $|\mathbf{Q}|$ is
large, and the magnetic form factor is low so that the spectrum is
dominated by the phonon scattering which is proportional to
$|\mathbf{Q}|^2$ or $|\mathbf{Q}|^4$.\cite{Squires} The
non-magnetic background for the low-angle scans was estimated by
scaling the high-angle intensity as $|\mathbf{Q}|^2$. The
background that is independent of $|\mathbf{Q}|$ was estimated
from the energy gain side of the spectrum at low temperatures. The
total background including phonon scattering is shown in
Fig.~\ref{Fig-background} as grey solid circles. Assuming that the
whole intensity at high scattering angles scales with
$|\mathbf{Q}|^2$ or that part of the scattering scales with
$|\mathbf{Q}|^4$ leads to a yet smaller background. The estimated
non-magnetic background scattering accounts only for a small
fraction of the observed scattering at $Q_c=1$ up to
$12\;\mathrm{meV}$ and substantiates the magnetic origin of the
observed continuum.\par

The momentum dependence of the continuum shows a broad maximum at
$Q_c=1$. Fig.~\ref{continuum-cut} compares two constant-$\omega$
scans for $\langle \omega \rangle=10\;\mathrm{meV}$ and for
$\langle \omega \rangle=14\;\mathrm{meV}$; the latter can be taken
as the background. The constant-$\omega$ data were obtained by
integrating the intensity between $8$ and $12\;\mathrm{meV}$ and
between $12$ and $16\;\mathrm{meV}$, respectively. Thus the
magnetic scattering from the well-defined excitation was not
included in the integration, while the results for $\langle \omega
\rangle=14\;\mathrm{meV}$ are a measure of the background.\par

\begin{figure}
\begin{center}
  \includegraphics[height=6.3cm,bbllx=55,bblly=262,bburx=482,
  bbury=572,angle=0,clip=]{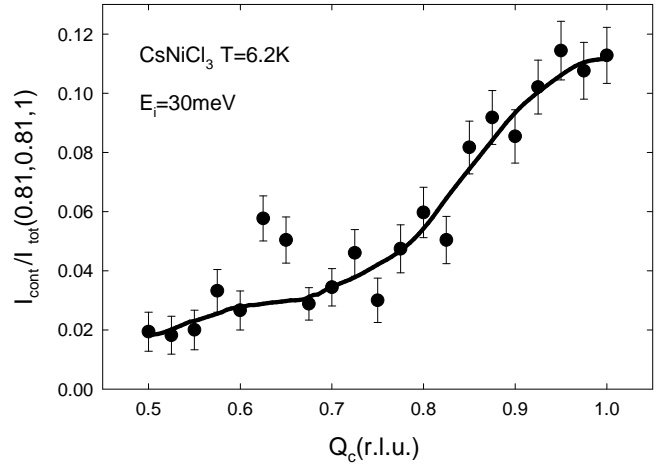}
  \vspace{0.1cm}
  \caption{Energy-integrated continuum scattering at
  $T=6.2\;\mathrm{K}$ as a fraction of the integrated intensity
  at $(0.81,0.81,1)$ and as a function of wave-vector transfer
  $Q_c$ along the chain. The data is corrected for the magnetic
  form factor. The solid line is a guide to the eye.}
  \label{continuumQ-fraction}
\end{center}
\end{figure}

Fig.~\ref{continuum-cut} shows that for wave-vector transfers $0.6
< Q_c < 1.4$, the scattering intensity for $\langle \omega
\rangle=10\;\mathrm{meV}$ is higher than for $\langle \omega
\rangle=14\;\mathrm{meV}$. The fact that the additional scattering
is located around the AF zone center is further evidence for its
magnetic origin.\par

The energy-integrated intensity of the multi-particle states for
$0.5 < Q_c < 1$ is shown in Fig.~\ref{continuumQ-fraction} as a
fraction of the total scattering intensity at $Q_c=1$ at the 1D
point $(0.81,0.81,1)$. It was determined numerically from the
measured intensity after the integrated intensity of the
well-defined excitation was subtracted.
Fig.~\ref{continuumQ-fraction} shows that the energy-integrated
intensity of the continuum has its maximum at $Q_c=1$ and
approaches zero for $Q_c=0.5$. The integrated intensity is smaller
near $Q_c=0.5$ partly because the range of energies for which the
multi-particle continuum can be observed gets smaller as the
momentum decreases.\par

\begin{figure}
\begin{center}
  \includegraphics[height=6.4cm,bbllx=60,bblly=260,bburx=485,
  bbury=570,angle=0,clip=]{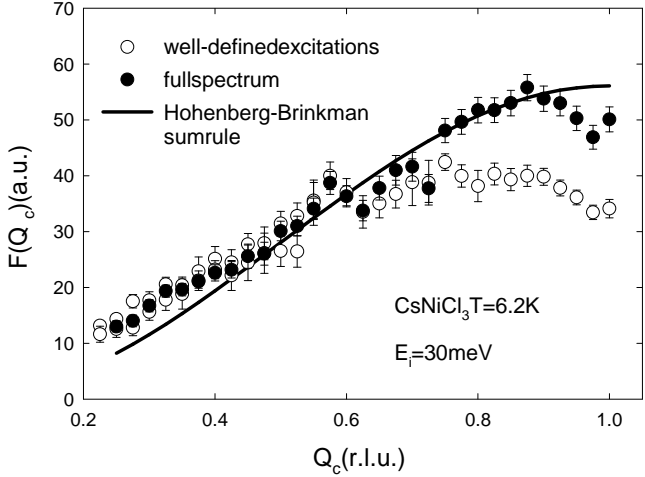}
  \vspace{0.1cm}
  \caption{$F(Q_c)=\int d\omega\, \omega\, S(Q,\omega)$ at
  $T=6.2\;\mathrm{K}$ as a function of the wave-vector transfer
  $Q_c$ along the chain. Solid circles show $F(Q)$ calculated
  numerically from the observed spectra, open circles represent
  $F(Q)$ of the well-defined excitations. The solid line is the
  result of the Hohenberg-Brinkman sum rule.}
  \label{fig-Hohenberg-Brinkman}
\end{center}
\end{figure}

\textit{Sum Rule:} A further test of the magnetic nature of the
continuum scattering is provided by the Hohenberg-Brinkman first
moment sum rule. The first energy moment $F(Q_c)$ = $\int
d\omega\, \omega\, S(Q,\omega)$ of 1D magnets with
nearest-neighbor exchange interaction is given by $F(Q_c) \propto
(1-\cos(\pi Q_c))$.\cite{Hohenberg_Brinkman} This relation is
valid even in the presence of a weak interchain coupling
constant.\cite{Kenzelmann_Santini} The first energy moment
$F(Q_c)$ was determined from the observed spectra after
subtracting a flat background and correcting for the magnetic form
factor. The contribution from the well-defined excitation is shown
in Fig.~\ref{fig-Hohenberg-Brinkman} as open circles and from the
total scattering by the solid circles. The solid line is the
prediction of the Hohenberg-Brinkman sum rule which predicts that
the first moment doubles between $Q_c=0.5$ and $Q_c=1$. It is
clear that the observed spectrum is consistent with the sum rule,
while $F(Q_c)$ determined from the well-defined excitations alone
is not. This result confirms that the multi-particle continuum is
magnetic, that the single-mode approximation fails and that the
presence of the continuum is required to fulfill the sum rule.\par

The presence of the high-energy continuum has important
consequences for the static structure factor $S(Q_c)$. Because
scattering at higher energies enters the first moment with a
higher weight than low-energy scattering, $S(Q_c)$ is smaller if
high-energy excitations are present. Since the continuum is most
intense for $Q_c=1$ and lower for increasing $|Q_c-1|$ the width
of $S(Q_c)$ as a function of $Q_c$ is expected to increase around
$Q_c=1$, corresponding to a reduced correlation length. This is
what we have indeed observed (see later).\par

\textit{Temperature Dependence:} The temperature dependence of the
multi-particle continuum was investigated between $6.2$ and
$12\;\mathrm{K}$. Fig.~\ref{continuum-6K-12K} compares the
scattering intensity at $Q_c=1$ for $6.2$ and $12\;\mathrm{K}$.
The two spectra were measured in the same configuration, with the
incident energy $E_i=30\;\mathrm{meV}$ and the c-axis
perpendicular to the incident neutron beam direction. At the
higher temperature, the well-defined excitation has a higher
energy and becomes wider, consistent with our previous triple-axis
measurements.\cite{Kenzelmann_CsNiCl3_gap} The model of the
antisymmetrized Lorentzian function weighted by the Bose factor
describes the well-defined excitations well, but the fits cannot
account for the scattering at higher energies
(Fig.~\ref{continuum-6K-12K}). At both $T=6.2$ and
$12\;\mathrm{K}$ the multi-particle scattering extends up to about
$12\;\mathrm{meV}$. Between $6.2$ and $12\;\mathrm{K}$ the
multi-particle scattering is largely independent of temperature
and the scattered intensity between $5$ and $12\;\mathrm{meV}$
energy transfer is the same at both temperatures. This result was
used earlier \cite{Kenzelmann_CsNiCl3_continuum} to average the
two data sets to produce a color plot with increased
statistics.\par

The upward renormalization of the excitation energy and the
broadening of the excitation with increasing temperature increases
the scattering at higher energy transfers as the temperature is
increased. Much above $12\;\mathrm{K}$ the multi-particle states
cannot be distinguished unambiguously from the broadened single
particle
excitation.\cite{Kenzelmann_CsNiCl3_gap,Kenzelmann_CsNiCl3_temperature}

\subsection{Correlation length}
\begin{figure}
\begin{center}
  \includegraphics[height=6.5cm,bbllx=70,bblly=260,bburx=482,
  bbury=570,angle=0,clip=]{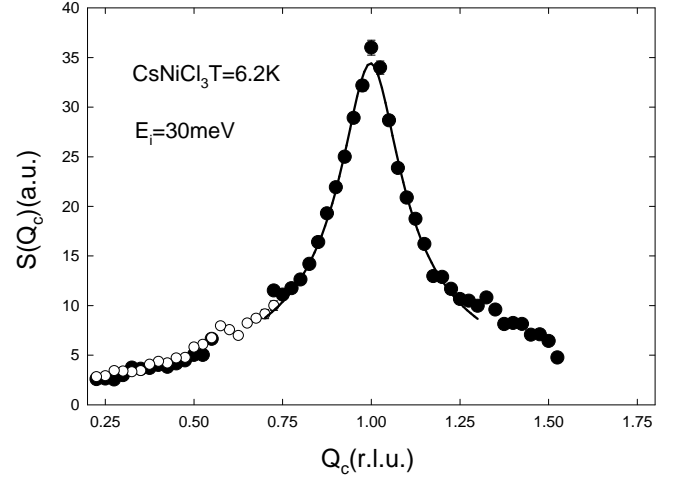}
  \vspace{0.2cm}
  \caption{Static from factor, $S(Q_c)$, for the wave-vector
  transfer $Q_c$ along the chain as a function of energy transfer
  at $T=6.2\;\mathrm{K}$. The open circles represent data measured
  at negative wave-vector transfer $Q_c$ and reflected at $Q_c=0$.
  The solid line is a fit to the data, based on Eq.~\ref{SRL-corrlength},
  as explained in the text.}
  \label{intensity-6K}
\end{center}
\end{figure}

The instantaneous structure factor at $T=6.2\;\mathrm{K}$ for the
1D chain, $S(Q_c)$, is shown in Fig.~\ref{intensity-6K}. It was
determined experimentally by integrating the scattering observed
with the MARI spectrometer over both the well-defined excitations
and the continuum scattering. The background scattering was
subtracted, the remaining scattering corrected for the magnetic
form factor dependence and then integrated numerically up to
energy transfers of $12\;\mathrm{meV}$ to give $S(\bbox{Q})$. In
the experiment the intensity was measured close to but not exactly
at the non-interacting wave-vector (inset in
Fig.~\ref{dispersion-width-6K}). The structure factor of the
independent chain was obtained from $S(\bbox{Q})$ by assuming as
for antiferromagnets that the integrated intensity of the
excitations close to the AF point is proportional to
$1/\omega(\bbox{Q})$ and the 1D structure factor is given by
\begin{equation}
    S(Q_c)=\frac{\omega(\bbox{Q})}{\omega(Q_c)}S(\bbox{Q})\, ,
    \label{3D-correction}
\end{equation}where $\omega(Q_c)$ and $\omega(\bbox{Q})$ are given
by Eqs.~\ref{1D-dispersion} and \ref{normalized_dispersion}.\par

If the correlations along the chains decay exponentially, as for a
1D Ising model, the structure factor is given by
\begin{equation}
    S_I(Q_c)=S^2\frac{\sinh(\xi_{I}^{-1})}{\cosh(\xi_{I}^{-1})+\cos(\pi Q_c)}\,
    .\label{Eq_static_classical}
\end{equation}where the correlation length $\xi_I$ is measured in
spin steps $c/2$ along the chain. If $\xi \gg 1$ and $|Q_c-1|\ll
1$ then this can be approximately written as a Lorentzian
\begin{equation}
    S_I(Q_c)= \frac{2\xi_I}{1+\pi^2(Q_c - 1)^2 \xi_{I}^2}\, .
    \label{Eq_Lorz_SQclassical}
\end{equation}When this function is convoluted with a Gaussian to
take account of the wave-vector range sampled along the chain
direction, and fitted to the $S(Q_c)$ shown in
Fig.~\ref{intensity-6K}, the correlation length obtained is
$2.2(2)$ spins. It would be only $1.9(1)$ spins if the 3D
correction, Eq.~\ref{3D-correction}, is neglected because the data
near $Q_c=1$ were taken for $Q_a<0.19$ where $S(\bbox{Q})$ is
suppressed.\par

The theory of the structure factor of $S=1$ linear chains is
usually formulated using the single-mode approximation in which it
is assumed that all the spectral weight resides in a mode whose
energy coincides with the first moment $\omega(Q_c)$ of the
Hohenberg-Brinkman sum rule:\cite{Hohenberg_Brinkman}
\begin{equation}
    S_{\rm SM}^{\alpha \alpha}(Q_c) = - \frac{4}{3}
    \frac{<\mathcal{H}>}{L} \frac{(1-\cos{\pi Q_c})}
    {\hbar \omega^{\alpha \alpha}(Q_c)}\, ,
    \label{SMA}
\end{equation}
where ${<\mathcal{H}>}/L$ is the ground state energy per spin.
Using Eq.~\ref{1D-dispersion} for $\omega(Q_c)$, $S(Q_c)$ near
$Q_c=1$ is proportional to a square-root Lorentzian
\begin{equation}
    S_{\rm SM}(Q_c) \propto \frac{\xi}{ \sqrt{ 1+ \pi^2 (Q_c-1)^{2}
    \cdot \xi^{2}}}\, ,
    \label{SRL}
\end{equation}where we expect
\begin{equation}
    \xi=\frac{\sqrt{v_{\rm s}^2+\alpha^2/4}}{\Delta}\, .
    \label{SRL-corrlength}
\end{equation}In Eq.~\ref{SRL} we neglect a small logarithmic term
necessary to conserve the total spin. When our experimental values
of $v_s$, $\alpha$ and $\Delta$ are inserted we expect
$\xi=4.8(2)$.\par

The square-root Lorentzian gives a better description of the data
in Fig.~\ref{intensity-6K} than the Lorentzian,
Eq.~\ref{Eq_Lorz_SQclassical}. After convoluting Eq.~\ref{SRL}
with a Gaussian to take account of the wave-vector range sampled
for each $Q_c$, and fitting to the data between $0.7 < Q_c < 1.3$,
we find $\xi=4.0(2)$ sites and if the 3D correction,
Eq.~\ref{3D-correction}, is neglected $3.0(2)$ sites. The value of
$4.0(2)$ less than $\xi=5$ from numerical calculations for
isolated chains.\cite{Kim} It is also less than the value of
$4.8(2)$ deduced from the parameters of the well-defined
excitations indicating that the single-mode approximation is not
entirely valid. The correlation length deduced from Eq.~\ref{SRL}
is shorter because of the presence of the continuum, which
depresses $S(Q_c)$ near $Q_c=1$ as may be seen from the first
moment sum rule. Qualitatively the continuum excitations at high
energy contribute less to $S(Q_c)$ than would a spectrum that was
entirely confined to low-energy excitations.\par

The observed value of $\xi$ at $6.2\;\mathrm{K}$ is much lower
than the result reported by Zaliznyak \textit{et
al.}\cite{Zaliznyak} This is because their measurements were
performed with wave-vectors close to the minimum of the excitation
energy at $\bbox{Q}_{\rm order}=(0.33,0.33,1)$. Consequently, the
results are strongly influenced by 3D effects because the
correlation length measured at this wave-vector diverges at the 3D
ordering temperature $T_N$. Those results cannot therefore be
compared with the correlation length of isolated quantum chains.
Furthermore Zaliznyak \textit{et al.}\cite{Zaliznyak} performed
the energy integration only up to energy transfers of
$1.6\;\mathrm{meV}$ and our results show that there is appreciable
scattering at higher energy transfers. This effect and their
chosen wave-vector transfer both give rise to a larger correlation
length that that observed in our experiment.

\section{Discussion}
With improved measurements of the gapped quantum-disordered spin
system ${\rm CsNiCl_3}$ at $6.2\;\mathrm{K}$ we have mapped out
the behavior of weakly-coupled spin-1 chains. We find (1) that the
Haldane excitations have a spin velocity
$v_s=5.70(7)\;\mathrm{meV}=2.5$ $J$ - in excellent agreement with
field theory\cite{Kveshchenko_Chubukov} and numerical
calculations,\cite{Takahashi89,Sorensen_3,Sorensen_2,Kim} (2) that
the single-chain correlation length is shorter than that of
isolated chains, and (3) that the observed multi-particle
continuum at the AF point carries about $12(2)\%$ of the total
scattering. This confirms previous experimental
results\cite{Kenzelmann_CsNiCl3_continuum}, that the continuum is
much larger than expected theoretically.

\subsection{Continuum}
Multi-particle continua for AF wave-vectors have been predicted
for $T=0\;\mathrm{K}$ in the framework of the ${\rm NL\sigma M}$
and numerical techniques. In the framework of the ${\rm NL\sigma
M}$ the continuum consists of three-particle states whose dynamic
structure factor has a energy gap of $3\Delta_0$ and a pronounced
maximum in its intensity at $\sim 6\Delta_0$ ($\Delta_0$ is the
Haldane gap energy at zero
temperature).\cite{Horton_Affleck,Essler} Its continuum integrated
intensity from $3\Delta_0$ up to $20\Delta_0$ is only a tiny
fraction, $1\%$, of the total scattering at $Q_c=1$, even if the
coupling of the magnetic chains are taken into account in a Random
Phase approximation (RPA).\cite{Essler}\par

The diagonalization of the quantum eigenstates of a spin-1 chain
with $N=20$ sites predicts that at $Q_c=1$ the Haldane excitation
carries $97\%$ of the total scattering,\cite{Takahashi94} leaving
no more than $3\%$ of the scattering weight to contribute to
continuum scattering. More precise results can be obtained using
the density matrix renormalization group method. These
calculations show that at $T=0\;\mathrm{K}$ the Haldane excitation
in a spin-1 chain with $N=256$ sites carries $97.6(1)\%$ of the
total scattering at $Q_c=1$,\cite{Kuehner_White} predicting thus
at least 2.4 times more scattering than the ${\rm NL\sigma M}$.
This discrepancy is not very surprising because the three-particle
states at higher energies involve excitations far away from the AF
wave-vector where the field theory is known not to give accurate
results.\par

Clearly none of the theories can explain our observed result of
$12\%$ continuum scattering. We conclude therefore that the
multi-particle continuum observed in ${\rm CsNiCl_3}$ is not in
accord with theoretical results for an isolated AF spin-1
Heisenberg chain with nearest-neighbor interactions at
$T=0\;\mathrm{K}$.\par

\subsection{Comparison to Majorana Fermion Theory}
Other models for AF spin-1 chains assume that the spins also
interact via biquadratic exchanges. Biquadratic spin interactions
are thought to be small in most materials, but they cannot be
ruled out \textit{a priori}. The Hamiltonian for these models can
be written as
\begin{equation}
    {\mathcal{H}}=J\sum_{i}^{\rm chain}\;{\bf{S}}_{\rm i}\cdot
    {\bf{S}}_{\rm i+1}+ b\;({\bf{S}}_{i}\cdot{\bf{S}}_{i+1})^{2}\, .
\end{equation}
This Hamiltonian can be exactly solved for $b=-1$ (Armenian point)
and for $b=1/3$ which is the valence-bond-solid (VBS). It has been
shown numerically that the energy gap of this Hamiltonian is
non-zero for $-1<b<1$ with a maximum at
$b=0.41$.\cite{Schollwoeck_Jolicoeur} Close to the Armenian point,
the spin system can be treated in the framework of Tsvelik's
Majorana fermion theory\cite{Tsvelik}, which is a low-energy field
theory for linear spin-1 chains. This theory predicts that the
multi-particle scattering states carry $17\%$ of the intensity at
$Q_c=1$ for $b$ close to $-1$.\cite{Essler} At the VBS point, the
excitation spectrum has an energy gap
$\Delta=0.664J$.\cite{Schollwoeck_Jolicoeur} Numerical
calculations predict that the integrated intensity of the
well-defined excitation at $Q_c=1$ carries $99\%$ of the total
scattering and multi-particle states are weaker by a factor of $2$
compared to the Heisenberg point of the model
($b=0$).\cite{Takahashi94}\par

In order to test whether the multi-particle continuum could be
caused by a strong biquadratic term contained in the Hamiltonian
of the Majorana theory, where $17\%$ of the scattering at $Q_c=1$
lies in the continuum, we have searched at $6\;\mathrm{K}$, where
the gap is $\Delta=1.24\;\mathrm{meV}$, for the onset of the
continuum at $3\Delta$ and its pronounced maximum\cite{Essler} at
$\sim 5\Delta$. Our resolution is more than adequate to observe a
dip between the one-particle peak at $\Delta$ and a maximum of the
continuum at $5\Delta=6.2\;\mathrm{meV}$, but instead the spectrum
shows a steadily decreasing intensity in this energy range
(Fig.~\ref{continuumQ-C5-highstatistics}, \ref{continuumQ-cuts}).
This may arise because the 3D interactions cause a change in the
energy of the gap with the in-plane wave-vector $Q_a$. At
$6.2\;\mathrm{K}$, the minimum gap energy is about
$0.4\;\mathrm{meV}$ \cite{Steiner} and so the minimum energy for
the onset of the continuum is $1.2\;\mathrm{meV}$ and comparable
with the 1D gap energy $\Delta$. This effect together with the
thermal broadening of the one-particle excitations may be
sufficient to fill in the gap between the one-particle and the
continuum contribution to the scattering.\par

More importantly, however, a strong biquadratic term ($b=1$) not
only increases the spectral weight of the multi-particle
continuum, it also reduces the energy
gap\cite{Schollwoeck_Jolicoeur} in the excitation spectrum to
zero. This is not in accordance with the facts. In ${\rm
CsNiCl_3}$ the single-chain gap $\Delta$ at $6.2\;\mathrm{K}$ is
higher than expected from the bilinear exchange $J$. This suggests
that biquadratic interactions are weak and that the observed
magnetic multi-particle continuum cannot be explained by the
Majorana fermion theory.\par

\subsection{Correlation length}
The single-chain correlation length $\xi$ has been determined to
be $\xi=4.0(2)$, which is considerably shorter than $\xi=5$
obtained from quantum Monte Carlo calculations for finite chains
at $T=0.25J\simeq 6.2\;\mathrm{K}$.\cite{Kim} Since the
measurements of $\xi$ were made close to the non-interacting
wave-vector, $Q_a=0.81$, they should give the correlation length
of isolated chains. Nevertheless the discrepancy may arise if the
RPA approximation, Eq.~\ref{normalized_dispersion}, does not
remove all of the 3D effects adequately.\par

The ground state of an isolated chain has a hidden string order.
It is composed of states such as $|+-+-0+00-0+>$ in which the
antiferromagnetic ordering is maintained across intervening zero
($S_z=0$) sites. Possibly this order is destroyed more effectively
by 3D interactions than is predicted by the RPA approximation.\par

The strong quantum fluctuations in $S=1$ linear chains are largely
due to the low spin and low dimensionality. The fluctuations may
further be enhanced, and the correlation length decreased, if
there are frustrated interactions. In ${\rm CsNiCl_3}$ there are
two possibilities for frustrated interactions. Firstly the chains
are arranged on a hexagonal lattice so there are bound to be
frustrated interactions between the chains arranged at the corners
of the triangle. A second possibility is that the inter-chain
interactions may be between not only spins in the same basal plane
but also between spins in the neighboring chains that are at $\pm
c/2$ along the chains.\cite{Kambe} If these interactions are also
strong and antiferromagnetic, as might be the case for
super-exchange interactions, then even the interactions between
pairs of chains are at least partially frustrated.\par

Further theoretical work is needed to understand whether these
multiply-frustrated interchain interactions are responsible for
the shortened correlation length as well as the associated upward
renormalization of the Haldane gap\cite{Kenzelmann_CsNiCl3_gap},
or whether the effects result from a failure of the RPA
approximation.\par

\subsection{Origin of continuum}
We first compare with a material with smaller 3D interactions.
${\rm Ni(C_{2}H_{8}N_{2})_{2}NO_{2}ClO_{4}}$ (NENP) is a AF spin-1
chain system in which the interchain couplings are not frustrated
and much smaller than in ${\rm CsNiCl_3}$, but a large single-site
anisotropy splits the three Haldane excitations into two distinct
excitations.\cite{Renard} For NENP at $0.3\;\mathrm{K}$ the ratio
of the first moment at the AF point and at the maximum of the
dispersion is $R=F(1)/F(0.5) \sim 1.8(3)$ for the weighted average
of the doublet and singlet excitations if data published by Ma
\textit{et al.} \cite{Ma} is used for the calculations, and this
is in agreement with the theoretical value
$2$.\cite{Hohenberg_Brinkman} Neglecting the multi-particle
continuum in ${\rm CsNiCl_3}$ this same calculation yields
$R=1.2(2)$. This shows that in contrast to ${\rm CsNiCl_3}$, the
sharp excitations in NENP are enough to fulfill the
Hohenberg-Brinkman sum rule and no magnetic weight is expected at
higher energy transfers as in ${\rm CsNiCl_3}$ at
$6.2\;\mathrm{K}$. Any continuum would be hard to observe even in
highly-deuterated NENP because of residual hydrogen scattering.
Because NENP has weaker 3D interactions, this comparison suggests
that interchain interactions may be responsible for generating the
continuum observed in ${\rm CsNiCl_3}$.\par

We speculate that the multi-particle continuum is generated by the
AF coupling of the spin-1 chains, and multiply frustrated
interchain interactions might be decisive for the size of the
effect. A shortened AF correlation length - due to an
frustration-induced enhancement of quantum fluctuations - has two
main effects on the dynamic structure factor at $Q_c=1$. Firstly,
it leads to an additional upward renormalization of the gap energy
which is proportional to $1/\xi$ (see Eq.~\ref{SRL-corrlength}).
Secondly, a decreased correlation length results in a reduced
structure factor $S(Q_c=1)$ because the total scattering must be
constant (Eq.~\ref{SRL}). Additionally there is strong evidence
that the absolute value of $S({\bf Q}=(0.81,0.81,1))$ at low
temperatures is smaller than that of isolated chains by at least
$1.6$.\cite{Kenzelmann_Santini} If only a well-defined mode
existed it would have to increase its energy at $Q_c=1$ by a
factor of $1.6$ to satisfy the first moment sum rule of Hohenberg
and Brinkman.\cite{Hohenberg_Brinkman} However, the Haldane energy
is only $20\%$ higher relative to an uncoupled chain and so the
increase of Haldane energy is insufficient to compensate for the
loss of intensity. The $only$ way for the system to comply with
the sum rule is to create a scattering continuum at high energies,
as we find it does. This is evidence that the observed continuum
in ${\rm CsNiCl_3}$ is correlated with an enhancement of quantum
fluctuations due to multiply-frustrated 3D spin correlations.\par

\subsection{Similarity to two-spinon continuum of spin-1/2 chains}
\begin{figure}
\begin{center}
  \includegraphics[height=7cm,bbllx=64,bblly=220,bburx=500,
  bbury=580,angle=0,clip=]{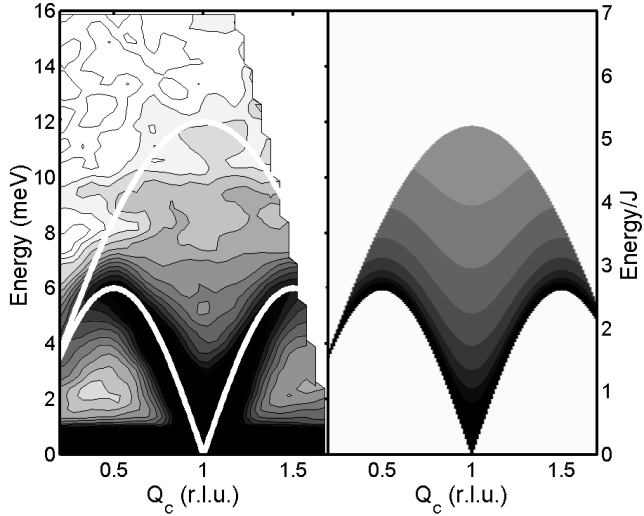}
  \vspace{0.2cm}
  \caption{Left panel: Scattering intensity observed in
  ${\rm CsNiCl_3}$ at $T=6.2\;\mathrm{K}$ as a function of
  wave-vector transfer $Q_c$ and energy transfer on a
  logarithmic scale. The data were measured with an incident
  energy $E_i=30\;\mathrm{meV}$. Data measured in
  detectors at high scattering angles are not shown.
  Right panel: Scattering intensity expected for a spin-1/2
  chain according to the M\"{u}ller Ansatz as a function of
  wave-vector transfer and energy transfer
  (in units of $J$).\protect\cite{Muller} The energy axis of
  the M\"{u}ller Ansatz was scaled so that the lower bound
  of the two-spinon continuum matches the maximum of the
  single-particle dispersion in ${\rm CsNiCl_3}$.}
  \label{fig-continuum-mueller-all}
\end{center}
\end{figure}

Fig.~\ref{fig-continuum-mueller-all} compares the observed
intensity at $6.2\;\mathrm{K}$ (left panel) with the scattering
expected for an AF spin-1/2 chain based on the M\"{u}ller Ansatz
\cite{Muller} (right panel). The peak of the lower energy boundary
of the two-spinon continuum was chosen to match the maximum of the
single-particle dispersion in ${\rm CsNiCl_3}$ for direct
comparison of the two excitation spectra. The experimentally
observed integrated intensity $I_{\rm tot}=\int dQ_c \int
d\omega\, S(Q_c,\omega)$ of ${\rm CsNiCl_3}$ with $S=1$ was set
equal to $S(S+1)=2$. The intensity of the M\"{u}ller Ansatz, which
is valid for chains with $S=1/2$, was scaled to $2S(S+1)=1.5$ (the
factor $2$ accounts for the spin-1/2 components per ${\rm
Ni^{2+}}$ site arising from the two holes in the d-shell of the
ion ${\rm Ni^{2+}}$).\par

The two excitation spectra in Fig.~\ref{fig-continuum-mueller-all}
have obvious similarities: the multi-particle continuum extends to
approximately $12\;\mathrm{meV}$, which is twice the maximum of
the one-particle energy, and the upper boundary of the continuum
decreases with increasing distance from the AF point. Both
continuum spectra can be described with a lower boundary
$\omega_{\rm min}=\omega_{\rm AFZB}|\sin(Q_c \pi)|$ (the
experiment could not establish whether the continuum scattering is
gapped or not) and an upper boundary $2 \omega_{\rm
max}=\omega_{\rm AFZB}\sin(\frac{Q_c \pi}{2})$, where $\omega_{\rm
AFZB} = 6\;\mathrm{meV}$. Qualitatively the continuum spectrum
found in ${\rm CsNiCl_3}$ has similar features to the continuum
found in AF $S=1/2$ chains.\cite{Tennant93}\par

\begin{figure}
\begin{center}
  \includegraphics[height=6.4cm,bbllx=70,bblly=262,bburx=481 ,
  bbury=572,angle=0,clip=]{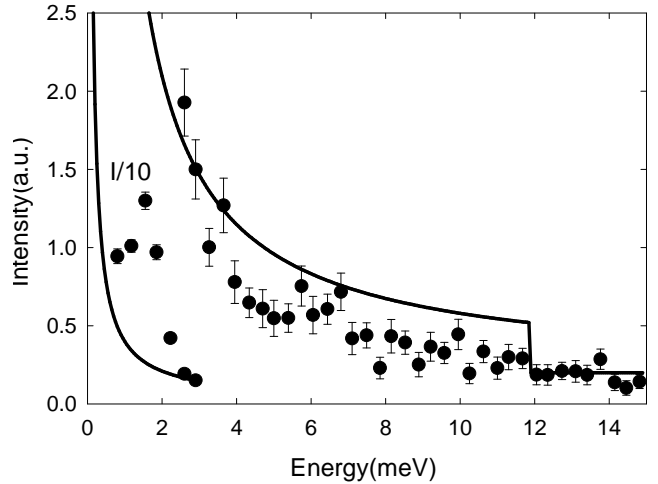}
  \vspace{0.2cm}
  \caption{Measured scattering intensity of ${\rm CsNiCl_3}$ at
  $\langle Q_c \rangle=1$ and at $T=6.2\;\mathrm{K}$ as a function
  of energy transfer. The solid line is the scattering intensity
  expected from a spin-1/2 chain according to the M\"{u}ller Ansatz.}
  \label{fig-continuum-mueller-pi}
\end{center}
\end{figure}

The intensity of the continuum in a spin-1/2 chain is much
stronger than that in ${\rm CsNiCl_3}$. This is shown in
Fig.~\ref{fig-continuum-mueller-pi}, which compares the scattering
at $Q_c=1$ as a function of energy transfer for ${\rm CsNiCl_3}$
and for a spin-1/2 chain. The lines of equal intensity are also
different in the two cases (Fig.~\ref{fig-continuum-mueller-all}).
At $7\;\mathrm{meV}$, as an example, the intensity in ${\rm
CsNiCl_3}$ is constant, while it has a clear minimum at $Q_c=1$
for the M\"{u}ller Ansatz.\par

We conclude that the intensity and the wave-vector dependence of
the observed continuum suggests that a small amount of spin-1/2
degrees of freedom are generated in ${\rm CsNiCl_3}$, possibly by
multiply-frustrated interchain interactions. In analogy to the
excitations in a spin-1/2 chain, the excitations in ${\rm
CsNiCl_3}$ may therefore also be described by a single spectral
function containing a sharp onset and a continuum at higher
energies. We found that the spectral functions given by field
theory \cite{Schulz} are indeed able to describe both the
well-defined excitation and the multi-particle continuum close to
the AF point if the critical exponents are adjusted, suggesting
that the well-defined excitation is an onset of a scattering
continuum. More theoretical work is needed to clarify the spectral
function and its dependence on the 3D interactions as the likely
source of the multi-particle states.\par

\section{Conclusions}
In summary, the excitation spectrum of ${\rm CsNiCl_{3}}$ has been
measured between $6.2$ and $12\;\mathrm{K}$. The results confirm
that the spin system is in a quantum-disordered phase and the
excitation spectrum has its full $Q_c=2$-periodicity. The
single-chain correlation length at $6.2\;\mathrm{K}$ is $4.0(2)$
sites along the chain, which is lower than the value predicted by
quantum Monte Carlo calculations, possibly due to spin
correlations perpendicular to the chain axis. The reduced
correlation length and the associated upward renormalization of
the Haldane gap underlines the need for more powerful theoretical
tools to describe the frustrated AF coupling of quantum
chains.\par

The key result is the observation of a magnetic multi-particle
continuum for wave-vectors $0.6 < Q_c < 1.4$ along the chain. The
magnetic multi-particle states have higher energies than the
well-defined excitations and extend up to about twice the maximum
in the one-particle dispersion. The integrated intensity of the
multi-particle continuum is much larger than that predicted for
the three-particle continuum by field theoretical models and
numerical techniques for AF spin-1 chains with nearest-neighbor
exchange interaction. The observed multi-particle continuum
resembles the two-spinon continuum of AF spin-1/2 chains, but its
intensity is weaker. This suggests the presence of spin-1/2
degrees of freedom in the quantum-disordered phase just above the
ordering temperature, possibly generated by competing interactions
due to the 3D AF interactions.\par

\begin{acknowledgments}
We would like to thank O. Petrenko (ISIS), S.~M. Bennington (ISIS)
and R.~L. Donaberger (Chalk River) for their assistance with
experiments and A. Zheludev (Oak Ridge National Laboratory), P.
Santini (Oxford University), F.~H.~L. Essler (Warwick University)
and I. Affleck (University of British Columbia) for enlightening
discussions. Financial support for the experiments was provided by
the EPSRC, by the EU through its Large Installations Program and
by the British Council-National Research Council Canada Program.
ORNL is managed for the U.S. D.O.E. by UT-Battelle, LLC, under
contract no. DE-AC05-00OR22725. One of the authors (M.~K.) was
supported by the Swiss National Science Foundation under Contract
No. 83EU-053223.
\end{acknowledgments}

\bibliographystyle{prsty}

\begin{thebibliography}{10}

\bibitem{Haldane83}
F.~D.~M. Haldane, Phys. Rev. Lett. {\bf 50},  1153  (1983).

\bibitem{Buyers86}
W.~J.~L. Buyers, R.~M. Morra, R.~L. Armstrong, M.~J. Hogan, P.
Gerlach, and K.
  Hirakawa, Phys. Rev. Lett. {\bf 56},  371  (1986).

\bibitem{Morra}
R.~M. Morra, W.~J.~L. Buyers, R.~L. Armstrong, and K. Hirakawa,
Phys. Rev. B
  {\bf 38},  543  (1988).

\bibitem{Ma}
S. Ma, C. Broholm, D.~H. Reich, B.~J. Sternlieb, and R.~W. Erwin,
Phys. Rev.
  Lett. {\bf 69},  3571  (1992).

\bibitem{Tennant93}
D.~A. Tennant, T.~G. Perring, R.~A. Cowley, and S.~E. Nagler,
Phys. Rev. Lett.
  {\bf 70},  4003  (1993).

\bibitem{Kenzelmann_CsNiCl3_continuum}
M. Kenzelmann, R.~A. Cowley, W.~J.~L. Buyers, R. Coldea, J.~S.
Gardner, M.
  Enderle, D.~F. McMorrow, and S.~M. Bennington, Phys. Rev. Lett. {\bf 87},
  017201  (2001).

\bibitem{Steiner}
M. Steiner, K.Kakurai, J.~K. Kjems, D. Petitgrand, and R. Pynn, J.
Appl. Phys.
  {\bf 61},  3953  (1987).

\bibitem{Zaliznyak}
I.~A. Zaliznyak, L.-P. Regnault, and D. Petitgrand, Phys. Rev. B
{\bf 50},
  15824  (1994).

\bibitem{Katori}
A.~A. Katori, Y. Ajiro, T. Asano, and T. Goto, J. Phys. Soc. Jpn.
{\bf 64},
  3038  (1995).

\bibitem{Kadowaki}
H. Kadowaki, K. Ubukoshi, and K. Hirakawa, J. Phys. Soc. Jpn. {\bf
56},  751
  (1987).

\bibitem{Cox_Minkiewicz}
D.~E. Cox and V.~J. Minkiewicz, Phys. Rev. B {\bf 4},  2209
(1971).

\bibitem{Yelon}
W.~B. Yelon and D.~E. Cox, Phys. Rev. B {\bf 7},  2024  (1973).

\bibitem{Minkiewicz}
V.~J. Minkiewicz, D.~E. Cox, and G. Shirane, Solid State Commun.
{\bf 8},  1001
   (1970).

\bibitem{Enderle99}
M. Enderle, Z. Tun, W.~J.~L. Buyers, and M. Steiner, Phys. Rev. B
{\bf 59},
  4235  (1999).

\bibitem{Affleck_Wellman}
I. Affleck and G.~F. Wellman, Phys. Rev. B {\bf 46},  8934
(1992).

\bibitem{Tun88}
Z. Tun, W.~J.~L. Buyers, R.~L. Armstrong, E.~D. Hallman, and D.~P.
Arovas, J.
  de Physique {\bf C1 49},  1431  (1988).

\bibitem{Affleck_Weston}
I. Affleck and R.~A. Weston, Phys. Rev. B {\bf 45},  4667  (1992).

\bibitem{Achiwa}
N. Achiwa, J. Phys. Soc. Jpn. {\bf 27},  561  (1969).

\bibitem{Moses}
D. Moses, H. Shechter, E. Ehrenfreund, and J. Makovsky, J. Phys. C
{\bf 10},
  433  (1977).

\bibitem{Kenzelmann_CsNiCl3_gap}
M. Kenzelmann, R.~A. Cowley, W.~J.~L. Buyers, and D.~F. McMorrow,
Phys. Rev. B
  {\bf 63},  134417  (2001).

\bibitem{Kenzelmann_CsNiCl3_temperature}
M. Kenzelmann, R.~A. Cowley, W.~J.~L. Buyers, R. Coldea, M.
Enderle, and D.~F.
  McMorrow, The temperature dependence of single particle excitations in a
  spin-1 chain from the quantum to the infinite-temperature limit, to be
  published.

\bibitem{Tun90}
Z. Tun, W.~J.~L. Buyers, R.~L. Armstrong, K. Hirakawa, and B.
Briat, Phys. Rev.
  B {\bf 42},  4677  (1990).

\bibitem{Jolicoeur_Golinelli}
T. ${\rm Jolic\oe ur}$ and O. Golinelli, Phys. Rev. B {\bf 50},
9265  (1994).

\bibitem{Sorensen_1}
E.~S. Sorensen and I. Affleck, Phys. Rev. B {\bf 49},  13235
(1994).

\bibitem{Sorensen_2}
E.~S. Sorensen and I. Affleck, Phys. Rev. B {\bf 49},  15771
(1994).

\bibitem{Zaliznyak_continuum}
I.~A. Zaliznyak, S.-H. Lee, and S.~V. Petrov, Phys. Rev. Lett.
{\bf 87},
  017202  (2001).

\bibitem{Cooper_Nathans}
M.~J. Cooper and R. Nathans, Acta Crys. {\bf 23},  357  (1967).

\bibitem{Squires}
G.~L. Squires, {\em Thermal Neutron Scattering} (Cambridge
University Press,
  Cambridge, 1978).

\bibitem{Hohenberg_Brinkman}
P.~C. Hohenberg and W.~F. Brinkman, Phys. Rev. B {\bf 10},  128
(1974).

\bibitem{Kenzelmann_Santini}
M. Kenzelmann and P. Santini, Temperature dependence of single
particle
  excitations in a ${\rm \textit{S}=1}$ chain: exact diagonalization
  calculations compared to neutron scattering experiments, submitted to Phys.
  Rev. B.

\bibitem{Kim}
Y.~J. Kim, M. Greven, U.~J. Wiese, and R.~J. Birgeneau, Eur. Phys.
J. B {\bf
  4},  291  (1998).

\bibitem{Kveshchenko_Chubukov}
D.~V. Kveshchenko and A.~V. Chubukov, Sov. Phys. JETP {\bf 66},
1088  (1987).

\bibitem{Takahashi89}
M. Takahashi, Phys. Rev. Lett. {\bf 62},  2313  (1989).

\bibitem{Sorensen_3}
E.~S. Sorensen and I. Affleck, Phys. Rev. Lett. {\bf 71},  1633
(1993).

\bibitem{Horton_Affleck}
M.~D.~P. Horton and I. Affleck, Phys. Rev. B {\bf 60},  11891
(1999).

\bibitem{Essler}
F.~H.~L. Essler, Phys. Rev. B {\bf 62},  3264  (2000).

\bibitem{Takahashi94}
M. Takahashi, Phys. Rev. B {\bf 50},  3045  (1994).

\bibitem{Kuehner_White}
T.~D. K\"{u}hner and S.~R. White, Phys. Rev. B {\bf 60},  335
(1999).

\bibitem{Schollwoeck_Jolicoeur}
U. Schollw\"{o}ck, T. ${\rm Jolic\oe ur}$, and T. Garel, Phys.
Rev. B {\bf 53},
   3304  (1996).

\bibitem{Tsvelik}
A.~M. Tsvelik, Phys. Rev. B {\bf 42},  10499  (1990).

\bibitem{Kambe}
T. Kambe, H. Tanaka, S. Kimura, H. Ohta, M. Motokawa, and K.
Nagata, J. Phys.
  Soc. Jpn. {\bf 65},  1799  (1996).

\bibitem{Renard}
J.~P. Renard, L.~P. Regnault, and M. Verdaguer, Journal de
Physique {\bf 49},
  1425  (1988).

\bibitem{Muller}
G. M\"{u}ller, H. Thomas, H. Beck, and J.~C. Bonner, Phys. Rev. B
{\bf 24},
  1429  (1981).

\bibitem{Schulz}
H.~J. Schulz, Phys. Rev. B {\bf 34},  6372  (1986).

\end{thebibliography}

\end{document}